\documentclass[aps,prl,twocolumn,superscriptaddress,oneside]{revtex4}
\usepackage{amsmath,graphicx,bm,textcomp}
\usepackage{placeins}
\usepackage{bbm}
\usepackage{pdfpages}
\usepackage{xcolor}
\graphicspath{{Figures/}}

\begin{document}
\title{Non-local signatures of the chiral magnetic effect in Dirac semimetal Bi$_{0.97}$Sb$_{0.03}$ }
	
\author{Jorrit C. de Boer}
\thanks{These two authors contributed equally.}
\affiliation{MESA$^+$ Institute for Nanotechnology, University of Twente, The Netherlands}
\author{Daan H. Wielens}
\thanks{These two authors contributed equally.}
\affiliation{MESA$^+$ Institute for Nanotechnology, University of Twente, The Netherlands}
\thanks{These two authors contributed equally.}
\author{Joris A. Voerman}
\affiliation{MESA$^+$ Institute for Nanotechnology, University of Twente, The Netherlands}
\author{Bob de Ronde}
\affiliation{MESA$^+$ Institute for Nanotechnology, University of Twente, The Netherlands}
\author{Yingkai Huang}
\affiliation{Van der Waals - Zeeman Institute, IoP, University of Amsterdam, The Netherlands}
\author{Mark S. Golden}
\affiliation{Van der Waals - Zeeman Institute, IoP, University of Amsterdam, The Netherlands}
\author{Chuan Li}
\affiliation{MESA$^+$ Institute for Nanotechnology, University of Twente, The Netherlands}
\author{Alexander Brinkman}
\affiliation{MESA$^+$ Institute for Nanotechnology, University of Twente, The Netherlands}

\today
\begin{abstract}
The field of topological materials science has recently been focussing on three-dimensional Dirac semimetals, which exhibit robust Dirac phases in the bulk. However, the absence of characteristic surface states in accidental Dirac semimetals (DSM) makes it difficult to experimentally verify claims about the topological nature using commonly used surface-sensitive techniques. The chiral magnetic effect (CME), which originates from the Weyl nodes, causes an $\textbf{E} \cdot \textbf{B}$-dependent chiral charge polarization, which manifests itself as negative magnetoresistance. We exploit the extended lifetime of the chirally polarized charge and study the CME through both local and non-local measurements in Hall bar structures fabricated from single crystalline flakes of the DSM Bi$_{0.97}$Sb$_{0.03}$. From the non-local measurement results we find a chiral charge relaxation time which is over one order of magnitude larger than the Drude transport lifetime, underlining the topological nature of Bi$_{0.97}$Sb$_{0.03}$.
\end{abstract}

\maketitle

\section{I. Introduction}
The electronic structure of bismuth-antimony alloys has been thoroughly studied in the 1960s and 70s and later regained attention when Bi$_{1-x}$Sb$_{x}$ was one of the first topological materials to be discovered \cite{Hsieh2008}. It was proposed that at the topological transition point, which is at $x\sim0.03$, an accidental band touching in the bulk of Bi$_{1-x}$Sb$_{x}$ makes this material a Dirac semimetal (DSM) \cite{Kim2013,Chuan2018,Nagaosa2014}. 
A popular method to measure the topological nature of DSMs is through the detection of the chiral magnetic effect (CME) in electronic transport measurements. However, these signals are often obscured by the presence of parallel, non-topological conduction channels and are, more importantly, difficult to distinguish from other effects that may cause negative magnetoresistance, such as current jetting \cite{Liang2018}. Parameswaran \emph{et~al}. \cite{Parameswaran2014} proposed to measure the CME non-locally, using the extended lifetime of chirally polarized charge. For the Dirac semimetal Cd$_3$As$_2$, where two sets of Dirac cones with opposite chirality are separated in momentum space and protected by inversion symmetry, this measurement technique has been succesfully used to measure the CME \cite{Zhang2017}. In this work, we present evidence for the presence of the chiral magnetic effect in the accidental Dirac semimetal Bi$_{0.97}$Sb$_{0.03}$, based on local and non-local magnetotransport results on crystalline, exfoliated flakes.

\begin{figure}
\includegraphics[width=.48\textwidth]{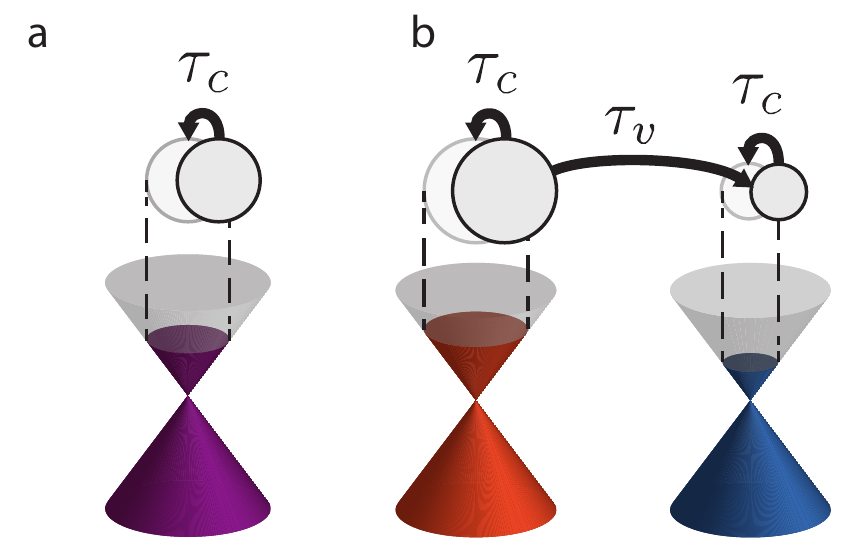}
\caption{\textbf{Dirac semimetals and the chiral magnetic effect.} (\textbf{a}) Two superposed Weyl cones in a Dirac semimetal. In an external electric field, the electrons on a shifted Fermi surface relax to their equilibrium state with characteristic time $\tau_c$. (\textbf{b}) In parallel external electric and magnetic fields, the two Weyl cones are separated in momentum space and exhibit different chemical potentials. Inter-cone relaxation of chirally polarized electrons occurs with characteristic time $\tau_{\text{\text{CME}}}$.}
\label{Fig:Fig1}
\end{figure} 

\begin{figure*}
\includegraphics[width=.96\textwidth,]{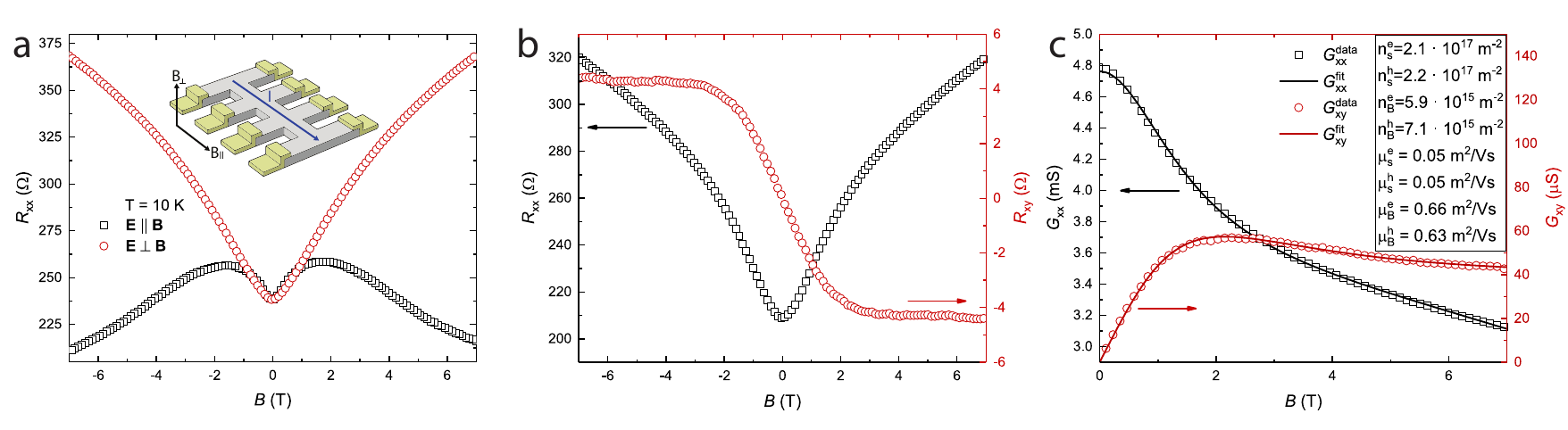}
\caption{\textbf{Local magnetoresistance.} (\textbf{a}) Local longitudinal magnetoresistance for perpendicular and parallel electric and magnetic fields. For parallel fields, the MR is strongly negative due to the chiral magnetic effect. Inset: schematic illustration of the device used for local transport. Current is sourced through 2 of the the 4 outer leads, while the inner 4 are used as probes for the longitudinal and Hall resistances, as is typical for Hall measurements. (\textbf{b}) Results of the Hall measurement for perpendicular electric and magnetic fields. (\textbf{c}) Drude multi-band fit on the conductance. The conductances have been obtained from the measured resistances $R_{\text{xx}}$ and $R_{\text{xy}}$ through tensor inversion. The low mobility, high carrier density surface state contributions are based on literature \cite{Chuan2018}.}
\label{Fig:Fig2}
\end{figure*} 

In a general Dirac semimetal, 2 Weyl cones of opposite chirality (often labeled as isospin degree of freedom \cite{Parameswaran2014}) are superposed in momentum space as depicted in figure \ref{Fig:Fig1}(a). Their chiralities are defined through an integral of the Berry connection, $A_{\textbf{k} \pm} = i \langle \psi_{\pm} | \nabla_{\textbf{k}} \psi_{\pm}\rangle$, over the Fermi surface (FS): 
\begin{equation}
	\chi_{\pm} = \frac{1}{2 \pi} \oint_{FS} (\nabla_{\textbf{k}} \times A_{\textbf{k} \pm}) \cdot d S_{\textbf{k}},
\end{equation}
where $\bm{k}$ is the wave vector. Using a basic representation of the Weyl nodes, $\psi_{\pm} = (e^{-i \theta} \sin(\phi/2), -\cos(\phi/2))^{T}$ (where $\theta$ is the in-plane angle in momentum space, with $\theta=0$ along the $k_x$-axis, and $\phi$ is the polar angle with respect to $k_z$), one finds $\chi_{\pm} = \pm 1$. These nodes with opposite chiralities always come in pairs and, when the degeneracy of the Weyl cones is lifted, they are connected in momentum space by a surface state known as a Fermi arc. In mirror symmetry-protected Dirac systems, such as Cd$_3$As$_2$, different pairs of Weyl nodes can also be connected by Fermi arcs, which has been experimentally observed \cite{Moll2016}. Bi$_{0.97}$Sb$_{0.03}$ contains accidental Dirac points at the 3 L-points \cite{Kim2013}. The crossings at these Dirac points are not protected by any symmetry, and should not be connected by Fermi arcs. However, upon breaking time reversal symmetry with an external magnetic field, the Dirac cones split into two Weyl cones of opposite chirality, in which case Bi$_{0.97}$Sb$_{0.03}$ behaves very similar to Cd$_3$As$_2$. A more in-depth discussion on the topological properties of Bi$_{0.97}$Sb$_{0.03}$ compared to those of Cd$_3$As$_2$ can be found in appendix A. 

The different chiral nodes correspond to a source and drain of Berry curvature ($\Omega_{\textbf{k} \pm} = \nabla_{\textbf{k}} \times A_{\textbf{k} \pm}$), which in turn behaves as a magnetic field in momentum space. Taking this momentum space analogue of the magnetic field into account in the equations of motion, one ends up with an expression for chiral charge pumping $\partial \rho_{\pm} / \partial t$ in external parallel electric and magnetic fields \cite{Nielsen1983,SonSpivak2013}:
\begin{equation}
	\frac{\partial \rho_{\pm}}{\partial t} = f(\Omega_{\textbf{k}}) \frac{e^3}{4 \pi^2 \hbar^2} \textbf{E} \cdot \textbf{B},
\end{equation}
where $\bm{E}$ and $\bm{B}$ are the electric and magnetic fields respectively, and $f(\Omega_{\textbf{k}})$ represents the chirality-dependence. The result from this semi-classical argument can also been obtained in the quantum limit \cite{Zyuzin2012}. This chiral charge pumping creates a difference in chemical potential in the two Weyl cones, causing a net imbalance in chirality, which eventually relaxes by means of impurity scattering. However, the orthogonality of the two degenerate Weyl cones with different isospin and the large momentum difference between different valleys suppress these relaxation processes. As a consequence, chiral charge has an increased lifetime compared to the Drude transport lifetime, which is shortened by many low-energy scattering events. Through the continuity equations, this chiral charge imbalance contributes to the longitudinal conductivity and can be observed in magnetotransport measurements  \cite{Kim2013,QLi2016}.

In Cd$_3$As$_2$, there are 2 scattering events that relax the chiral charge polarization: scattering between degenerate cones of different isospin, and intervalley scattering. Of these two, intervalley scattering should be expected to be the dominant factor as the momentum difference between the valleys is relatively small \cite{Zhang2017}. In Bi$_{0.97}$Sb$_{0.03}$, where the Dirac points reside at the 3 L-points, intervalley scattering requires a momentum transfer of the order $2\pi/a$, with $a$ being the lattice constant. Because of this required large momentum transfer, we argue that isospin-flip scattering through multiple scattering events is the likely dominant chiral relaxation process in Bi$_{0.97}$Sb$_{0.03}$. 

\section{II. Methods and characterization}

To characterize the Bi$_{0.97}$Sb$_{0.03}$ crystals (which are grown as described by Li \textit{et al.} \cite{Chuan2018}), several devices with contacts in a Hall bar configuration were fabricated. For all devices in this work, flakes of Bi$_{0.97}$Sb$_{0.03}$ were exfoliated from single crystals onto SiO$_2$/Si$^{++}$ substrates. Contact leads were defined using standard e-beam lithography, followed by sputter deposition of 120 nm Nb with a few nm of Pd as capping layer, and lift-off. Then, the flakes themselves were structured using another e-beam lithography step, now followed by Ar$^+$ milling. Magnetotransport measurements were conducted at 10~K in He-4 cryostats.

\begin{figure}
\includegraphics[width=.48\textwidth]{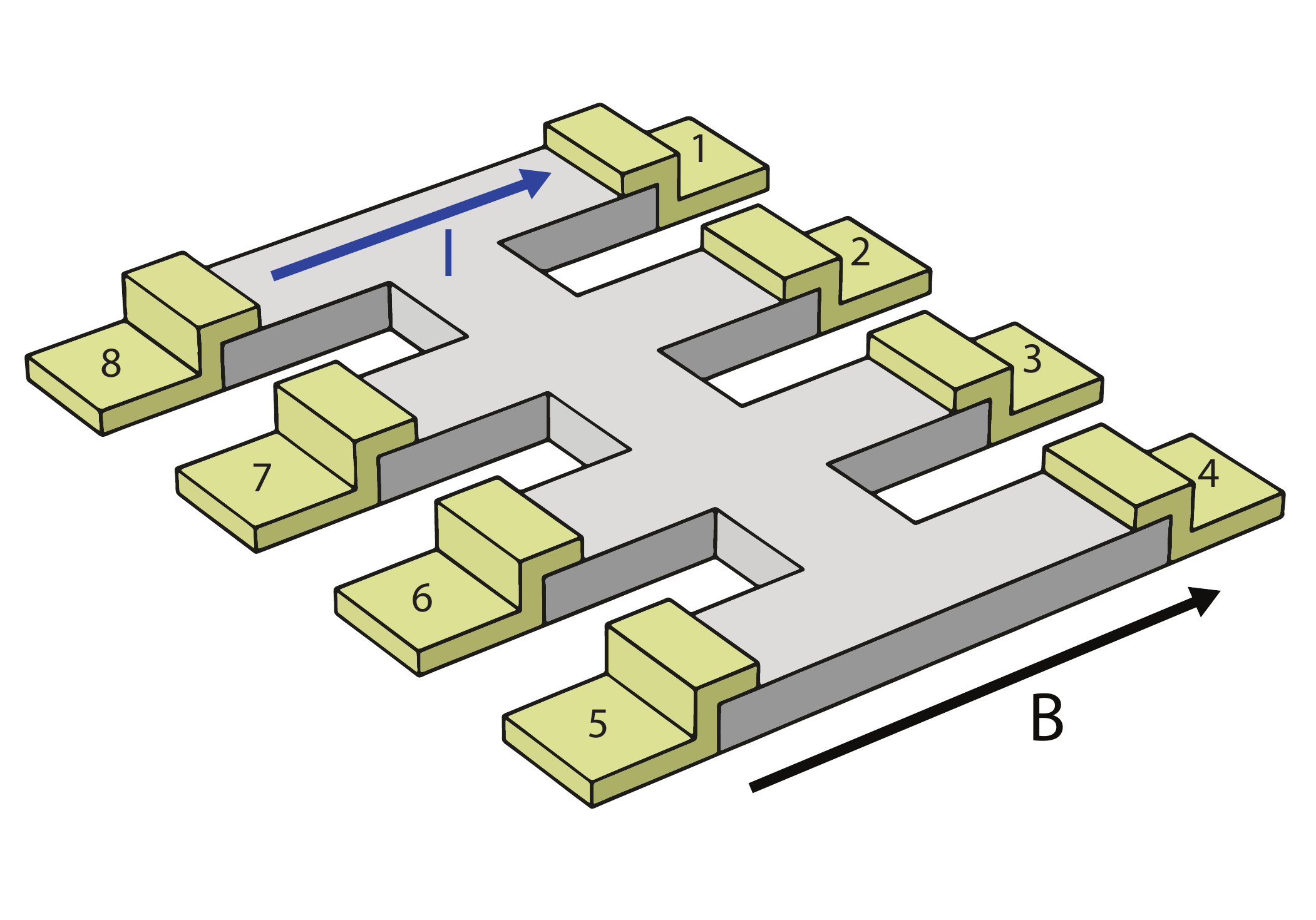}
\caption{\textbf{Non-local setup.} A schematic illustration of the device used for non-local transport measurements (gray represents the Bi$_{0.97}$Sb$_{0.03}$ flake and yellow the metallic contacts). Current is sourced parallel to the external magnetic field, through 2 contacts on the side, and the chiral polarization that diffuses into the central channel, is measured at the contacts further along this channel.}
\label{Fig:Fig3}
\end{figure} 


The inset of figure \ref{Fig:Fig2}(a) shows a schematic overview of the measurement setup as used for local magnetotransport measurements. A current is sourced through the outer contacts and voltages are measured at the contacts in between. 
As observed earlier by Kim \textit{et al.} \cite{Kim2013}, the magnetoresistance of Bi$_{0.97}$Sb$_{0.03}$, shown in figure \ref{Fig:Fig2}(a), exhibits negative magnetoresistance for parallel electric and magnetic fields. This negative magnetoresistance is considered to be an indication of the CME \cite{Kim2013,HuiLi2013,Caizhen2015,QLi2016}. While the CME in Bi$_{0.97}$Sb$_{0.03}$ originates from the bulk electrons, the magnetoresistance data shows no Shubnikov-de Haas (SdH) oscillations corresponding to the bulk electron pockets, despite the low effective mass and high mobility of these electrons \cite{LiuAllen1995}. We will comment on this later. However, in accordance with previous measurements on flakes of Bi$_{0.97}$Sb$_{0.03}$, we do observe SdH oscillations originating from the bulk hole pocket in different samples made of the same single crystal (see appendix B).

Figure \ref{Fig:Fig2}(b) shows the results of a Hall-type measurement. By tensor inversion of the measured longitudinal and Hall resistances, the longitudinal and transverse conductances were obtained. In figure~\ref{Fig:Fig2}(c), the conductances are fitted using a multi-band model, which takes two surface and two bulk conduction channels into account~\cite{Chuan2018}. For the bulk electrons, we obtain a bulk electron density of $n_B^e =$ 3.0 $\cdot 10^{22}$ m$^{-3}$, where we have used a flake thickness of 200 nm. For anisotropic Fermi velocities of $v_1=$ 0.8$\cdot 10^5$~m/s and $v_2=$ 10$\cdot 10^5$~m/s \cite{Hsieh2008}, this would indicate that the Fermi energy lies only $E_F = \hbar(\pi^2 n_B^e e v_1 v_2 /3)^{1/3} = $ 13 meV above the Dirac point. The bulk charge carrier mobilities as obtained from the multi-band fit are lower than those found in unstructured devices \cite{Chuan2018}. This is in line with the absence of SdH oscillations in this measurement, which can be attributed to the device dimensions being of the same order as the cyclotron radius, and to disorder due to etching at the device edges. The consequential broadening of the Landau levels does not hamper the presence of the CME \cite{Zhang2017}. For a conservative effective mass of $m_{e,h} = 0.05 \, m_0$ \cite{Chuan2018}, the bulk electron and hole  mobilities of $\mu = $0.65~m$^2$/Vs give us an estimate of the momentum relaxation time: $\tau_c = \mu m /e \approx 1.8 \cdot 10^{-13}$~s. 
	This is in line with the absence of SdH oscillations in this measurement, which can be attributed to the device dimensions being of the same order as the cyclotron radius.

\begin{figure*}
\includegraphics[width=.96\textwidth]{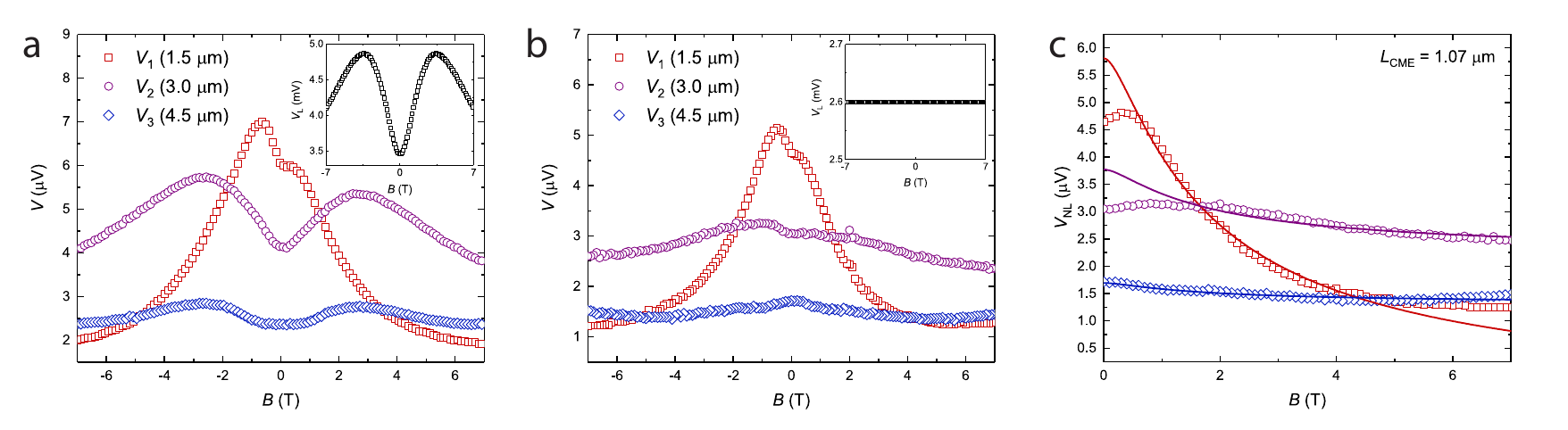}
\caption{\textbf{Non-local signals.} (\textbf{a}) Measured voltages at distances 1.5, 3 and 4.5 $\mu$m. The inset shows the local voltage, measured at the source contacts. (\textbf{b}) Same as in panel (a), but for a varying source current $I_s(B)$ such that $V_{\text{L}}=\text{cst.}$ (inset). (\textbf{c}) Symmetrized non-local voltages for fixed local electric field, $V_{\text{NL}}(E=\text{cst.\,},B)$, from panel (b).}
\label{Fig:Fig4}
\end{figure*} 

\begin{figure}[b]
\includegraphics[width=.48\textwidth]{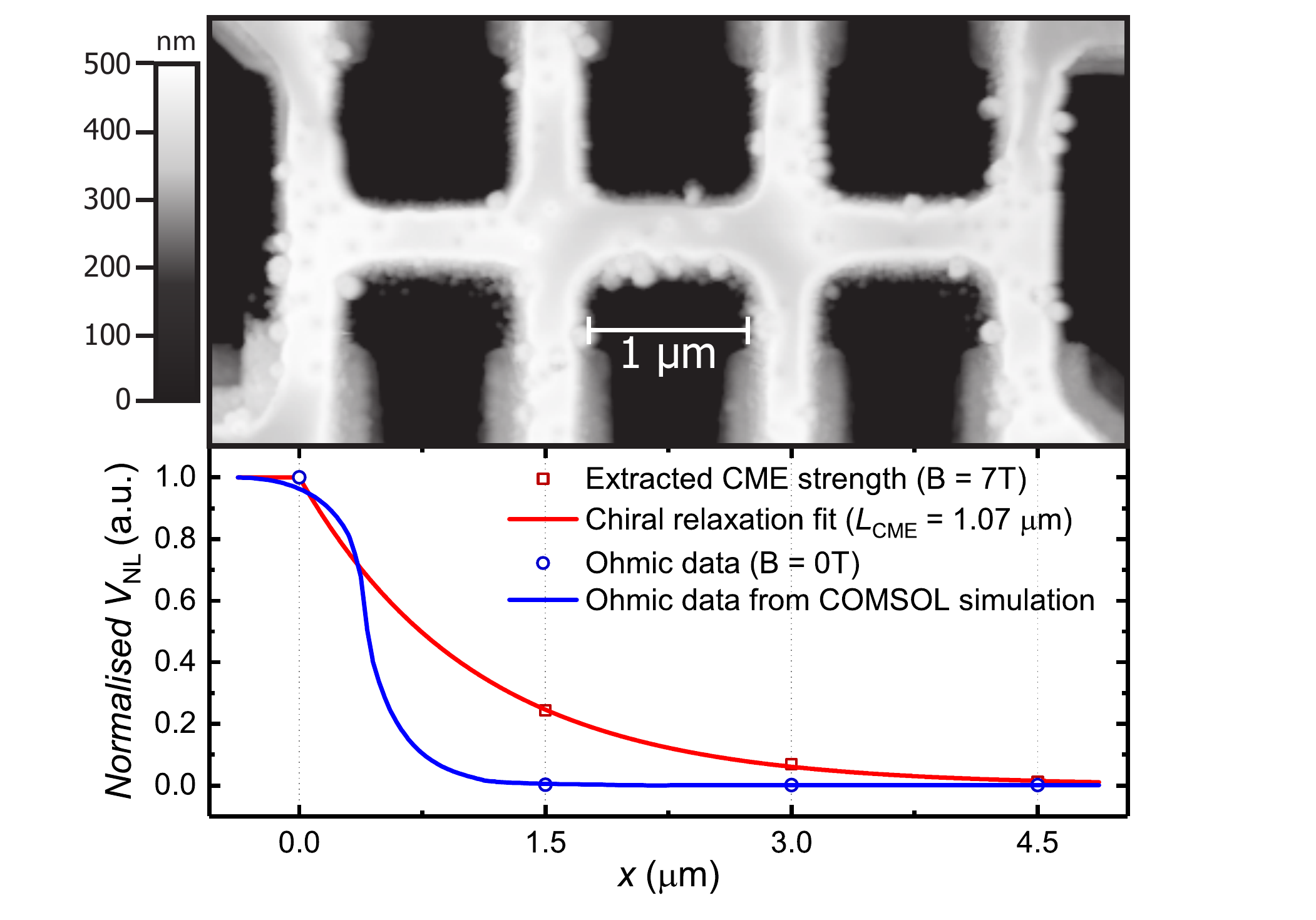}
\caption{\textbf{Diffusion lengths.} Upper panel: Atomic force microscopy image of a Bi$_{0.97}$Sb$_{0.03}$ single crystal flake, structured into a device for non-local measurements and contacted with Nb leads. Lower panel: Normalized strength of the Ohmic (zero field) and CME (from fit) contributions to the signals measured at the voltage probes. The CME strength persists over longer distances than the Ohmic signal.}
\label{Fig:Fig5}
\end{figure} 

\begin{figure*}[t]
\includegraphics[width=.96\textwidth]{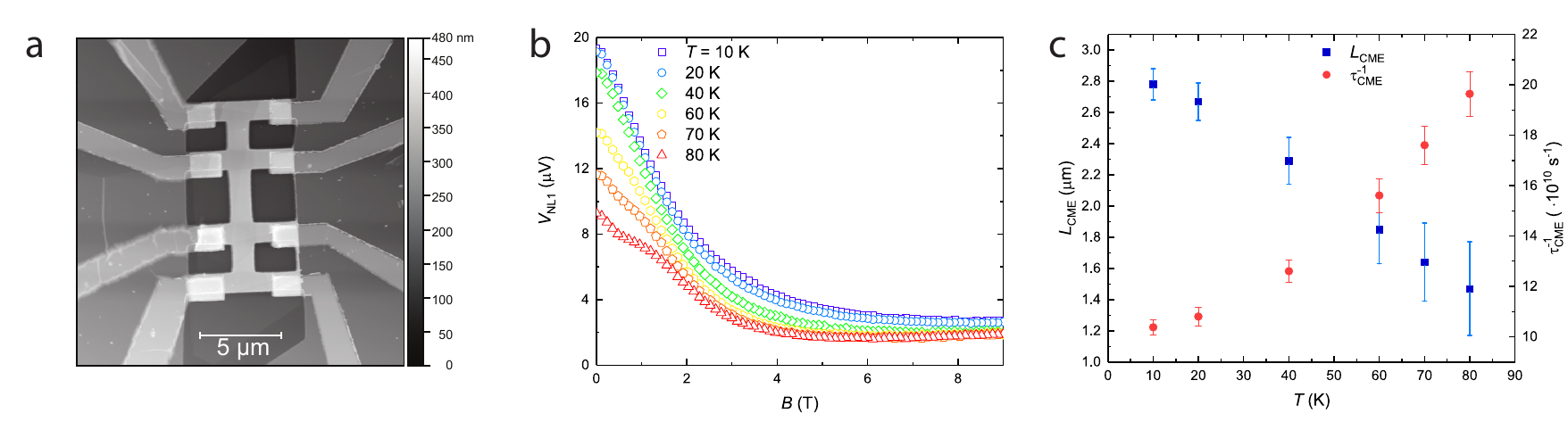}
\caption{\textbf{Temperature dependence of the CME.} (\textbf{a}) Atomic force microscopy image of the device used for non-local measurements of the CME on a Bi$_{0.97}$Sb$_{0.03}$ flake at different temperatures. (\textbf{b}) Temperature dependence of the voltage at the non-local contact closest to the source. (\textbf{c}) Temperature dependence of the chiral charge diffusion length $L_{\text{\text{CME}}}$ and the chiral polarization lifetime $\tau_{\text{\text{CME}}}$ as extracted through Eqn. (\ref{eqn:VNL}). For more information, see appendix F.}
\label{Fig:Fig6}
\end{figure*} 

The non-local measurement setup is designed such that we can measure the coupling between the polarization of chirality and an external magnetic field at different distances from the polarization source, and is shown in figure \ref{Fig:Fig3}. A current is sourced from contact 8 to 1, as indicated by the blue arrow. By applying a magnetic field parallel to the current, a chiral charge imbalance is induced.  As the charge diffuses away from the polarizing source, the polarization becomes weaker and so does the measurable voltage of the polarized charge in the external magnetic field. We measure the voltages locally ($V_L\equiv V_{81}=V_8-V_1$) and non-locally ($V_1\equiv V_{72}$, $V_2\equiv V_{63}$ and $V_3\equiv V_{54}$). To be able to distinguish the Ohmic (i.e. normal diffusion) and CME signals, the voltage terminals are located at distances similar to both the expected Ohmic and chiral relaxation lengths. 

When studying the CME, in the ideal case one measures the non-local response of the chiral anomaly as a function of the applied magnetic field only, i.e. keeping the applied electric field at the source contacts constant. However, due to the low resistance of these samples, we cannot voltage bias our sample and must resort to a current source, thereby causing the current to be constant as a function of the applied magnetic field. When measuring the local electric field, we find that this field is dependent on the magnetic field as well.

In order to obtain a data set with a constant electric field at the source contacts, measurements were performed by sweeping the source current from 0 to 10~$\mu$A for every magnetic field point, resulting in local and non-local magnetoresistance curves for a range of source currents $V_{\text{L,1,2,3}}(I_s,B)$. An example of such a set of magnetoresistance curves for a fixed source current is shown in figure \ref{Fig:Fig4}(a). The local voltage, $V_{\text{L}}(I_s,B)$, is shown in the inset and should be kept constant, which is achieved by varying the source current $I_s(V_{\text{L}}=\text{cst.\,},B)$. In figure \ref{Fig:Fig4}(b), we present the resulting voltages $V_{\text{NL}}(I_s(B),B)$, when the local electric field is kept constant in this way. Note that the local voltage, shown in the inset of figure \ref{Fig:Fig4}(b), is now constant. Using this method, the magnetic field dependence of the non-local signals can be studied without side effects from the local magnetoresistance. For more information on this procedure, see appendix C.

\section{III. Results}
The measured local voltage $V_{\text{L}}$, presented in the inset of figure \ref{Fig:Fig4}(a), is in good agreement with the expected resistance based on the 4-point resistance, taking the size of the channels into account. This indicates that the effects of contact resistances are negligible. Figure \ref{Fig:Fig4}(b) shows the non-local voltages measured at different distances as a function of the applied magnetic field. Here, the electric field at the source side is kept constant. At zero magnetic field, the measured voltages drop with increasing distance from the source. Furthermore, at all distances we observe a decreasing voltage with increasing magnetic field, which we attribute to the CME. The CME is not dominant for the entire magnetic field range as both at low and high fields, the voltage increases slightly with magnetic field. Kim \textit{et al.} attribute the low field MR to weak anti-localization \cite{Kim2013}. High field deviations from the CME signal may originate from higher order terms, which are not taken into account in this work. 

We have identified three possible causes of the small asymmetry of the data. First and foremost is the device asymmetry with respect to the source channel, but variations in sample thickness and imperfections in the structuring process may also be contributing factors. 
Figure \ref{Fig:Fig4}(c) shows the symmetrized non-local voltages. We fitted the intermediate field data between 2~T and 5.5~T with a model that subtracts the constant Ohmic contribution, and extracts the diffusion of the chiral charge as given by Parameswaran \textit{et al.} \cite{Parameswaran2014}:
\begin{equation}
V_{\text{NL}}(x)=-\left(\frac{B}{\gamma+B}\right)^2e^{-|x|/L_{\text{\text{CME}}}} + V_{\text{Ohmic}}.
\label{eqn:VNL}
\end{equation}
Here $x$ is the distance between the source and the non-local probes, $\gamma$ is proportional to the conductance at the metal contact and $L_{\text{\text{CME}}}$ is the diffusion length of the chiral charge polarization. The fit agrees well with the data for intermediate magnetic fields and it gives a diffusion length of 1.07~$\mu$m. Using $L_{\text{\text{CME}}} = \sqrt{D \, \tau_{\text{\text{CME}}}}$ and $D=(1/3)\,v_F\,l_m$, with $l_m = v_F \, \tau_{\text{\text{CME}}}$, we find a chiral polarization lifetime of $\tau_{\text{\text{CME}}} \approx 3.7 \cdot 10^{-12}$~s, which is over one order of magnitude longer than the Drude transport lifetime $\tau_c \approx 1.4 \cdot 10^{-13}$~s. 

The dependence of the normalized Ohmic and CME contributions to the measured voltages is presented in figure \ref{Fig:Fig5}, along with an atomic force microscopy (AFM) image of the device. Here, the amplitudes of the best fits are used to represent the CME strength. The measured Ohmic (zero-field) contribution at all voltage terminals is shown for comparison. The Ohmic contribution of the device is also modeled numerically, where the shown solid curve is a line cut along the horizontal part of the device (see appendix Dl). The simulated Ohmic contributions fit very well to the measured data, emphasizing the good homogeneity of the flake. The most notable feature of figure \ref{Fig:Fig5} is that the chiral polarization of the charge carriers has a significantly longer relaxation length than the Drude transport lifetime.

To study the temperature dependence of the CME in Bi$_{0.97}$Sb$_{0.03}$, another device was measured at higher temperatures. This device has larger channel widths and spacing as can be seen in the AFM image of the device in figure \ref{Fig:Fig6}(a). Figure \ref{Fig:Fig6}(b) shows the measured voltages at the non-local contacts closest to the source, which present the most striking features. It is apparent that for all temperatures displayed in this figure, the magnetoresistance is strongly negative and that this non-local voltage decreases as temperature increases. Through the same fitting procedure described above, the chiral charge polarizarion diffusion length is extracted for each temperature and shown in figure \ref{Fig:Fig6}(c) (more information on this procedure can be found in appendix F). In contrast to what has been found for Cd$_3$As$_2$ \cite{Zhang2017}, the chiral diffusion length in Bi$_{0.97}$Sb$_{0.03}$ does not seem to be constant with increasing temperature, but rather decreases linearly. The increasing relaxation rate $\tau_{\text{\text{CME}}}^{-1}$ with increasing temperature, indicates that inelastic processes are responsible for the relaxation of the chiral polarization in Bi$_{0.97}$Sb$_{0.03}$.

\section{IV. Conclusions}
In summary, we studied the chiral magnetic effect in Bi$_{0.97}$Sb$_{0.03}$ through transport measurements in local and non-local configurations. First, we characterized the exfoliated crystalline Bi$_{0.97}$Sb$_{0.03}$ flakes using a Hall-type measurement. Here, we identified contributions from two bulk bands, one of them corresponding to the electron pockets with a linear dispersion and a Fermi level close to the Dirac point. When subjected to parallel electric and magnetic fields, local measurements on our Bi$_{0.97}$Sb$_{0.03}$ devices show a pronounced negative magnetoresistance, an indication of the chiral magnetic effect. 

In a non-local configuration, we measured voltages that strongly decrease with increasing magnetic field, which we attribute to the chiral magnetic effect. As voltage contacts are located further away from the polarization source, the measured chiral magnetic effect weakens. This weakening occurs at a much lower rate than the decay of the Ohmic signal, which is a consequence of the long lifetime of the chiral polarization $\tau_{\text{CME}}$. Furthermore, measurements at different temperatures show that the chiral charge diffusion length decreases with increasing temperature, emphasizing the role of inelastic scattering in the chiral charge relaxation process in Bi$_{0.97}$Sb$_{0.03}$. Both local and non-local measurements provide strong evidence of the presence of the chiral magnetic effect in the three-dimensional Dirac semimetal Bi$_{0.97}$Sb$_{0.03}$.

\begin{acknowledgements}
\section{Acknowledgements}
The authors would like to thank T. Hashimoto for fruitful discussions and acknowledge financial support from the European Research Council (ERC) through a Consolidator Grant and from the Netherlands Organization for Scientific Research (NWO) through a Vici Grant.
\end{acknowledgements}


\clearpage

\renewcommand{\theequation}{S\arabic{equation}}
\renewcommand{\thefigure}{S\arabic{figure}}
\renewcommand{\bibnumfmt}[1]{[S#1]}
\renewcommand{\citenumfont}[1]{S#1}
\renewcommand{\thetable}{S\arabic{table}}
\graphicspath{{Figures/}}

\appendix
\onecolumngrid
\begin{center}
  \textbf{\large Supplemental Material to: Non-local signatures of the chiral magnetic effect in Dirac semimetal Bi$_{0.97}$Sb$_{0.03}$}\\[.2cm]
  Jorrit C. de Boer,$^{1,*}$ Daan H. Wielens,${1,*}$, Joris A. Voerman,$^1$ Bob de Ronde,$^1$\\ Yingkai Huang,$^1$ Mark S. Golden,$^2$ Chuan Li$^1$ and Alexander Brinkman$^1$\\[.1cm]
  {\itshape ${}^1$MESA$^+$ Institute for Nanotechnology, University of Twente, The Netherlands\\
  ${}^2$Van der Waals - Zeeman Institute, IoP, University of Amsterdam, The Netherlands}\\
\end{center}
\twocolumngrid

\setcounter{figure}{0}
\setcounter{equation}{0}
\section{Appendix A: Effective Hamiltonian and the chiral magnetic effect}
\subsection{A1. C\lowercase{d}$_3$A\lowercase{s}$_2$}

Cd$_3$As$_2$ is a Dirac semimetal belonging to the point group $D_{\text{\text{4h}}}$. The 2-fold inversion symmetry in this system ensures two topologically protected Dirac points along the $k_z$-axis. To explore the electronic structure of Cd$_3$As$_2$ we utilize a linearized version of the $k \cdot p$ model Hamiltonian as defined by B.J. Yang and Nagaosa \cite{SNagaosa2014}:
\begin{equation}
	H_{\text{\text{eff}}} =  v_\parallel(k_x \, \sigma_x s_z + k_y \, \sigma_y s_0) \pm v_z \, k_z \, \sigma_z s_0,
\end{equation}
where $\pm$ is used to switch between the nodes residing at $k_z' = \pm k_D$ and $k_z$ is measured relative to the center of the node. The total angular momentum and orbital degrees of freedom are indicated by $s$ and $\sigma$, respectively. We can see that Cd$_3$As$_2$ exhibits orbital-momentum locking and is of the form $ (\boldsymbol{k} \cdot \boldsymbol{\sigma}) s_0 $. Assuming isotropic spin-orbit coupling strength ($v_\parallel = v_z = v_F$) and switching to spherical coordinates with $\theta$ the angle in the $xy$-plane and $\varphi$ the polar angle measured from $+k_z$, we get 
\begin{equation}
	H_{\text{\text{eff}}} = 	\hbar v_F \begin{bmatrix} 
				\pm \cos{\varphi} &\sin{\varphi} \, e^{i \theta} &0 &0\\
				\sin{\varphi} \, e^{-i \theta} &\mp \cos{\varphi} &0 &0\\
				0 &0 &\pm \cos{\varphi} &-\sin{\varphi} \, e^{-i \theta}\\
				0 &0 &-\sin{\varphi} \, e^{i \theta} &\mp \cos{\varphi}
				\end{bmatrix},	
\label{eq:H_3}					
\end{equation} 
where the basis is taken as $\{ |s \uparrow \rangle, |p_x + i p_y\uparrow \rangle, |s \downarrow \rangle, |p_x - i p_y\downarrow \rangle \}$. In this case, the eigenenergies are simply $E_\pm = \pm v_F \textbf{\textit{k}}$. For a system doped to the n-type regime so that $E > 0$, we can find the normalized spinors at the $k_z = -k_D$ Dirac point, which we will refer to as node 1:
\begin{align}
\begin{split}
	\psi_+^1(\textbf{\textit{k}}) &= \big( e^{i \theta}\cos{\varphi/2}, \, \sin{\varphi/2}, \,0, \,0 \big)^{T}\\
	\psi_-^1(\textbf{\textit{k}}) &= \big( \,0, \,0, \,-e^{-i \theta}\cos{\varphi/2}, \, \sin{\varphi/2} \big)^{T}.
	\end{split}
\end{align} 
For node 2, located at $k_z = + k_D$, we have
\begin{align}
\begin{split}
	\psi_+^2(\textbf{\textit{k}}) &= \big( \sin{\varphi/2} \, e^{i \theta}, \, \cos{\varphi/2}, \,0, \,0 \big)^{T}\\
	\psi_-^2(\textbf{\textit{k}}) &= \big( \,0, \,0, \,-\sin{\varphi/2} \, e^{i \theta}, \, \cos{\varphi/2} \big)^{T}.
\end{split}
\end{align} 

\begin{figure*}
\includegraphics[width=.96\textwidth]{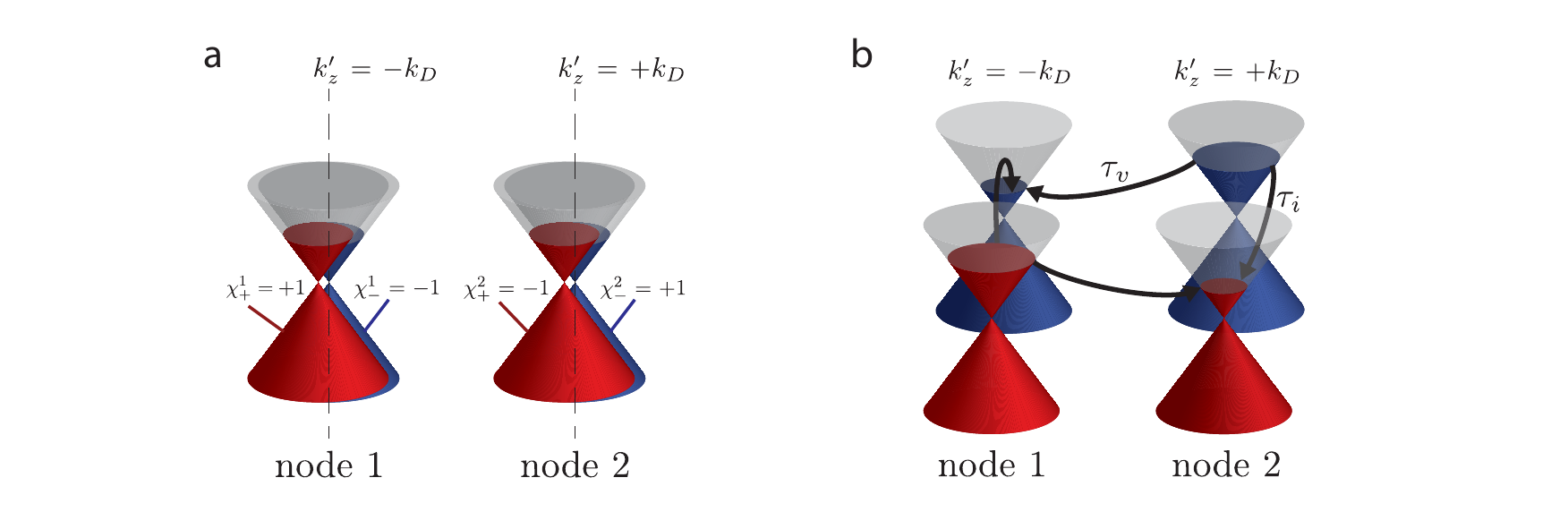}
\caption{\textbf{Schematic of the Cd$_3$As$_2$ band structure.} \textbf{(a)} Cd$_3$As$_2$ has 2 nodes (or `valleys') on the $k_z$-axis, each 2-fold degenerate ($\chi = \pm 1$). In valley 2, the chiralities are opposite to those in valley 1. \textbf{(b)} The chiral magnetic effect causes charge pumping from Fermi surfaces with $\chi = -1$ to those with $\chi = +1$. At a certain charge pumping rate, the `chiral charge imbalance' is equilibrated by intercone relaxation processes, which can be intervalley or isospin-flip processes. For clarity, the degenerate cones of different isospin were given an offset in panel (b).}
\label{fig:figA1}
\end{figure*}

To find the topological properties of node 1, we first need to find the Berry connection $A_+^1 = i \langle \psi_+^1(\textbf{\textit{k}}) | \nabla_k \psi_+^1(\textbf{\textit{k}}) \rangle$. Because in spherical coordinates $\nabla_k = \frac{\partial}{\partial k} \hat{k} + \frac{1}{k} \frac{\partial}{\partial \varphi} \hat{\varphi} + \frac{1}{k \sin{\varphi}} \frac{\partial}{\partial \theta} \hat{\theta}$, we get
\begin{equation}
	A_+^1 = -\frac{\cos^2{\varphi/2}}{k \, \sin{\varphi}} \hat{\theta}.	
\label{eq:Connection}					
\end{equation} 
From this, we find that the Berry curvature
\begin{equation}
	\Omega_+^1 = \nabla \times A_+^1 = \frac{1}{k \sin{\varphi}} \frac{\partial}{\partial \varphi} (A_+^1 \sin{\varphi})\hat{k} = \frac{1}{2 k^2} \hat{k}.	
\label{eq:Curvature}					
\end{equation} 
This Berry curvature can be seen as a magnetic field in $k$-space and indicates the monopole character of the node. The flux coming from each node gives us, once normalized by $2 \pi$, the ``chirality" or ``Chern number" of the node: 
\begin{equation}
	\chi_+^1 = \frac{1}{2\pi} \int_0^{2 \pi} \int_0^{\pi} \Omega_+^1 \, k^2 \sin{\varphi} \; d\varphi \, d\theta  = 1.	
\label{eq:Chirality}					
\end{equation} 
Similarly, we find $\chi_-^1 = -1$ and for node 2 we find $\chi_+^2 = -1$ and $\chi_-^2 = +1$. First of all, this shows us that the Dirac points in this material have Chern number $\pm 1$, which makes Cd$_3$As$_2$ topological. Furthermore, it seems that the degenerate Dirac cones have opposite chirality, and that the nodes at $k_z \rightarrow \mp k_D$ also have opposite chiralities (figure~\ref{fig:figA1}(a)), reflecting the 2-fold inversion symmetry of the system.

As the $k$-space analogue, the Berry curvature also has to be taken into account in the equations of motion that describe the equilibrium state of the system. Since the chiralities tell us that some Dirac points act as sources of Berry curvature and others as drains, the signs of the forces due to the Berry curvature are also opposite. This results in a net imbalance between cones of opposite chirality, a so called ``chiral charge imbalance". In this section, we will derive the resulting charge pumping by studying the Landau levels.

As Eqn. (\ref{eq:H_3}) consists of 2 decoupled Weyl cones of the form $H=\pm v_F\bm{\sigma}\cdot\bm{p}$, where $\bm{\sigma}$ denotes the orbital degree of freedom, one can easily find the Landau level dispersion relations as described by Zyuzin \textit{et al.} \cite{SZyuzin2012}. In short, one includes the vector potential correction to the momentum as $\bm{\pi}=\bm{p}+e\bm{A}$ and rewrites the Hamiltonian in terms of $a = (\pi_x - i \pi_y)/\sqrt{2 \hbar e B}$ and $a^{\dagger} = (\pi_x + i \pi_y)/\sqrt{2 \hbar e B}$ (for a magnetic field $\bm{B} = B_z$):
\begin{align}
\begin{split}
H &=\pm \hbar \omega (\sigma_+ a +\sigma_- a^{\dagger}) \pm \hbar v_F\sigma_z k_z \\
H &= \pm \begin{bmatrix} 
\hbar v_F k_z &\hbar \omega a \\
\hbar \omega a^{\dagger} &-\hbar v_F k_z 
\end{bmatrix},
\end{split}
\end{align} 
where $\sigma_{\pm} = (\sigma_x \pm i \sigma_y)/2$. Assuming a wavefunction of the form $ \psi_n = ( u_n |n-1\rangle, v_n |n\rangle)^T$, the creation and annihilation operators can be replaced according to $a|n\rangle = \sqrt{n} |n-1\rangle$ and $a^{\dagger}|n-1\rangle = \sqrt{n} |n\rangle$. To find the dispersion relation one can solve $\det{(H - \sigma_0 E_n)}=0$ with
\begin{equation}
H = \begin{bmatrix} 
\hbar v_F k_z  &\hbar \omega \sqrt{n}\\
\hbar \omega \sqrt{n} & -\hbar v_F k_z 
\end{bmatrix}
\end{equation}
to find $E_n = \pm \hbar \omega \sqrt{n + (v_F k_z /\omega)^2}$. Here, $\pm$ corresponds to the chirality of the accompanying wavefunction, as shown in Eqn. \ref{eq:Chirality}. For $n=0$, this gives us the dispersion of the zeroth Landau level: $E_0 = \pm \hbar v_F k_z$, which describes a linear dispersion with a Fermi velocity parallel or anti-parallel to $\bm{B} = B_z$, depending on the chirality of the Weyl node. It can be shown in a rather easy way that a dispersion relation of this form leads to a chiral charge imbalance \cite{SZyuzin2012}. The resulting charge pumping between Dirac cones of opposite chirality is balanced by relaxation processes (see figure~\ref{fig:figA1}(b)). Orthogonality of isospin and the large $\Delta \textbf{\textit{k}}$ involved with intervalley scattering significantly increase the relaxation times ($\tau_i$ and $\tau_v$ respectively), so that the chiral charge imbalance becomes a relevant quantity in transport measurements. 

\subsection{A2. B\lowercase{i}$_{1-x}$S\lowercase{b}$_x$}
Bi$_{1-x}$Sb$_x$ belongs to the point group $D_{\text{\text{3D}}}$ and has 3-fold rotation symmetry. There are no symmetries in this system that ensure the presence of Dirac cones, and the Dirac cones that do reside at the L-points are therefore labelled as accidental band touchings. Due to the 3-fold rotation symmetry, Bi$_{1-x}$Sb$_x$ has three accidental Dirac cones, separated by 120$^\circ$.
Topological Bi compounds can generally be described using the model Hamiltonian for topological insulators as developed by Liu \textit{et al.} \cite{SLiu2010}. Assuming isotropic Fermi velocities and taking $m \rightarrow 0$, the linearized Hamiltonian can be written as
\begin{equation}
H_{\text{\text{Liu}}} = \hbar v_F \sigma_x(s_x k_y - s_y k_x) + \hbar v_F  \sigma_y s_0 k_z.
\end{equation}
Teo \textit{et al.} \cite{STeo2008} described the Dirac physics around a single L-point in Bi$_{1-x}$Sb$_x$ in great detail using a modified form of this Hamiltonian, which can be obtained as $H_{\text{\text{TFK}}} = U_{\text{\text{R}}}^{\dagger} H_{\text{\text{Liu}}} U_{\text{\text{R}}}$ with $U_{\text{\text{R}}} = (s_0-i s_x-i s_y-i s_z)/2$ (a $-2 \pi/3$ rotation along the [111]-axis in spin space) and taking $v_x\rightarrow -v_x$:
\begin{equation}
H_{\text{\text{TFK}}} = \hbar v_F \sigma_x(s_x k_x + s_z k_y) + \hbar v_F \sigma_y s_0 \, k_z.
\end{equation}
The spinor part of the two wavefunctions for the conduction band side of the cone can be written as
\begin{align}
\begin{split}
	\psi_+(\textbf{\textit{k}}) &= \frac{1}{\sqrt{2}}\big( \cos{\varphi/2}, -e^{i \theta}\sin{\varphi/2}, e^{i \theta}\sin{\varphi/2},  \cos{\varphi/2}\big)^{T}\\
	\psi_-(\textbf{\textit{k}}) &= \frac{1}{\sqrt{2}}\big( \sin{\varphi/2}, e^{i \theta}\cos{\varphi/2}, e^{i \theta}\cos{\varphi/2},  -\sin{\varphi/2}\big)^{T}.
\end{split}	
\end{align} 
In the same manner as for Cd$_3$As$_2$, we can use these spinors to find the Berry curvature $\Omega_+^1 = \nabla \times A_{\pm} = \pm \frac{1}{2 k^2} \hat{k}$ and the chirality $\chi_{\pm} = \pm 1$. The unitary transformation $U_{\text{\text{R}}}$ can be used to transform $\psi_{\text{\text{TFK}}}$ into $\psi_{\text{\text{Liu}}}$, so that the topological properties of $H_{\text{\text{Liu}}}$ are the same as those found for $H_{\text{\text{TFK}}}$. To find out of this leads to the chiral magnetic effect in Bi$_{1-x}$Sb$_x$, we consider the full 4x4 Hamiltonian. 

For a magnetic field along the $c$-axis we use $H_{\text{\text{Liu}}}$, with the spin space rotated by 90$^\circ$ along the $s_z$-axis so that, in the basis $\{ |\sigma_1 \uparrow \rangle, |\sigma_1 \downarrow \rangle, |\sigma_2 \uparrow \rangle, |\sigma_2 \downarrow \rangle \}$, the effective Hamiltonian takes the form:
\begin{align}
\begin{split}
	H_{\text{\text{Liu}}}' &= \hbar v_F \sigma_x(s_x  k_x + s_y  k_y) + \hbar v_F \sigma_y s_0 \, k_z	\\
	&= \hbar v_F \begin{bmatrix} 
				0			&0 			&-i k_z		&k_x-i k_y\\
				0	 		&0 			&k_x+i k_y 	&-i k_z\\
				i k_z		&k_x-i k_y 	&0 			&0\\
				k_x+i k_y 	&i k_z		&0 			&0\\
				\end{bmatrix}.	
				\end{split}			
\end{align}
The orbital shift due to the vector potential is included as $\bm{\pi}=\bm{p}+e\bm{A}$, which we write again as creation and annihilation operators: $a = (\pi_x - i \pi_y)/\sqrt{2 \hbar e B}$ and $a^{\dagger} = (\pi_x + i \pi_y)/\sqrt{2 \hbar e B}$. The corresponding raising and lowering matrices are $s_{\pm} = \sigma_x(s_x \pm i s_y)$. Then
\begin{equation}
	H_{\text{\text{Liu}}}' = \begin{bmatrix} 
				0							&0 					&-i \hbar v_F k_z			&\hbar \omega a\\
				0	 						&0 					&\hbar \omega  a^{\dagger} 	&-i \hbar v_F k_z\\
				i \hbar v_F k_z				&\hbar \omega  a 	&0 							&0\\
				\hbar \omega  a^{\dagger} 	&i \hbar v_F k_z	&0 							&0\\
				\end{bmatrix}	,		
\end{equation}
where $\hbar \omega = v \sqrt{2 \hbar e B}$. As a trial wavefunction, we double the basis used for the 2x2 case: $ \psi_n = ( u_n^1 |n-1\rangle, v_n^1 |n\rangle, u_n^2 |n-1\rangle, v_n^2 |n\rangle)^T$. To find the dispersion relations for this wavefunction, we solve
\begin{equation}
	\det{\begin{bmatrix} 
				-E_n					&0 						&-i \hbar v_F k_z		&\hbar \omega \sqrt{n}\\
				0	 					&-E_n 					&\hbar \omega \sqrt{n} 	&-i \hbar v_F k_z\\
				i \hbar v_F k_z			&\hbar \omega \sqrt{n} 	&-E_n 					&0\\
				\hbar \omega \sqrt{n}	&i \hbar v_F k_z		&0 						&-E_n\\
				\end{bmatrix}}	= 0.		
\end{equation}
This gives us $E_n = \pm \hbar \omega \sqrt{n + (v_F k_z/\omega)^2}$, exactly the same result as we found for Cd$_3$As$_2$, including the linear zeroth Landau levels.

The simplified Cd$_3$As$_2$ Hamiltonian we employed earlier has isotropic orbital-momentum locking $ \hbar v_F \bm{\sigma} \cdot \bm{k} $, so that the Landau level formation is the same for magnetic fields in all directions. $H_{\text{\text{Liu}}}' = \hbar v_F \sigma_x(s_x  k_x + s_y  k_y) + \hbar v_F \sigma_y s_0 \, k_z$ has a clear difference between the spin-momentum locking in the in-plane $k_x$ and $k_y$ directions, and the $k_z$ direction, so a different response to magnetic field from different directions can be expected. To study the effect of a magnetic field along the $k_x$-axis ($B_x$), we first perform a rotation along the [111]-axis in spin space ($U_{\text{\text{R}}} = (s_0-i s_x-i s_y-i s_z)/2$) to get:
\begin{align}
\begin{split}
	H_{\text{\text{Liu}}}'' &= U_{\text{\text{R}}} H_{\text{\text{Liu}}}' U_{\text{\text{R}}}^{\dagger}\\
	&=\hbar v_F \sigma_x(s_y  k_x + s_z  k_y) + \hbar v_F \sigma_y s_0 \, k_z	\\
	&= \hbar v_F \begin{bmatrix} 
				0				&0 				&ky -i k_z		&-i k_x\\
				0	 			&0 				&i k_x 			&-(ky + i k_z)\\
				ky + i k_z		&-i k_x 		&0 				&0\\
				i k_x 			&- (ky-i k_z)	&0 				&0\\
				\end{bmatrix},								
\end{split}
\end{align}
which makes the following operations easier. With the modified raising and lowering matrices $s_{\pm} = \sigma_x s_z \pm i \sigma_y s_0)$ and operators $a = (\pi_y - i \pi_z)/\sqrt{2 \hbar e B}$ and $a^{\dagger} = (\pi_y + i \pi_z)/\sqrt{2 \hbar e B}$, we can rewrite $H_{\text{\text{Liu}}}''$ into
\begin{equation}
	H_{Liu}'' = \begin{bmatrix} 
				0							&0 					&\hbar \omega a 	&-i \hbar v_F k_x\\
				0	 						&0 					&i \hbar v_F k_x	&-\hbar \omega  a^{\dagger}\\
				\hbar \omega  a^{\dagger}	&-i \hbar v_F k_x	&0 					&0\\
				i \hbar v_F k_x				&-\hbar \omega  a	&0 					&0\\
				\end{bmatrix}.		
\end{equation}
With $ \psi_n = ( u_n^1 |n-1\rangle, v_n^1 |n\rangle, v_n^2 |n\rangle, u_n^2 |n-1\rangle)^T$, we find from
\begin{equation}
	\det{\begin{bmatrix} 
				-E_n					&0 							&\hbar \omega \sqrt{n} 	&-i \hbar v_F k_x\\
				0	 					&-E_n 						&i \hbar v_F k_x		&-\hbar \omega  \sqrt{n}\\
				\hbar \omega \sqrt{n}	&-i \hbar v_F k_x			&-E_n 					&0\\
				i \hbar v_F k_x			&-\hbar \omega  \sqrt{n}	&0 						&-E_n\\
				\end{bmatrix}}	= 0,		
\end{equation}
that the zeroth Landau level disperses as $E_0 = \pm \hbar v_F k_x$, which is same linear dispersion as for the $B_z$ field.

For a magnetic field in the $k_y$-direction ($B_y$), we can rotate $H_{\text{\text{Liu}}}$ around the [111]-axis in spin space in the different direction to get $H_{\text{\text{Liu}}}''' = U_R^{\dagger} H_{Liu}' U_R =\hbar v_F \sigma_x(s_z  k_x + s_x  k_y) + \hbar v_F \sigma_y s_0 \, k_z$, which gives a result analogous to the $B_x$ case: $E_0 = \pm \hbar v_F k_y$. This shows that the chiral zeroth Landau levels, and therefore also the CME, are expected to occur in every direction in Bi$_{1-x}$Sb$_x$.

\section{Appendix B: Local transport}
\begin{figure*}
\includegraphics[width=.96\textwidth]{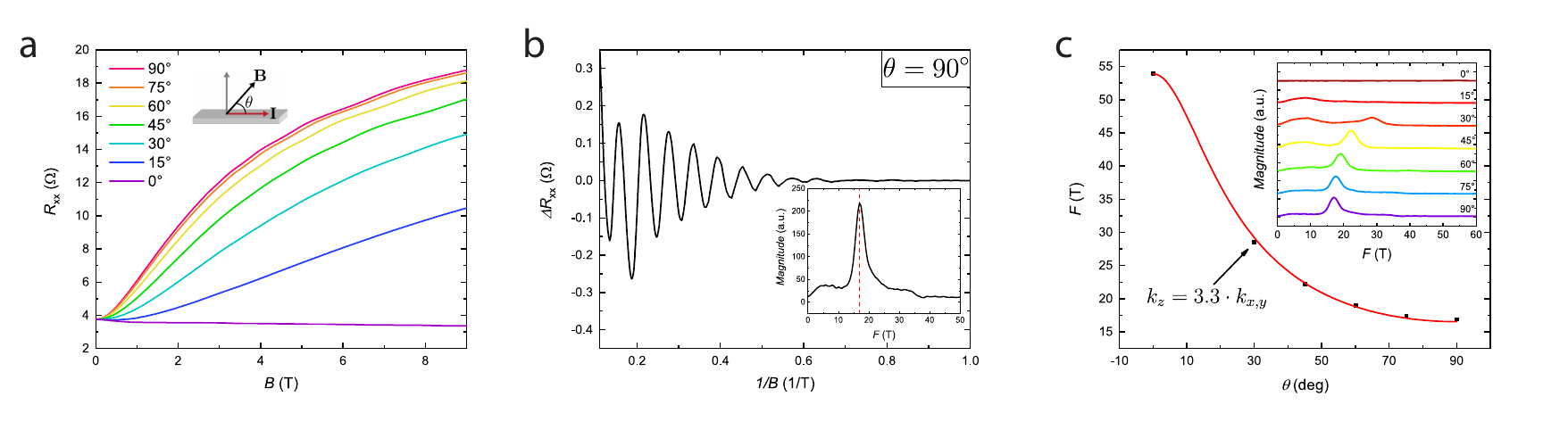}
\caption{\textbf{Local transport measurements on a Hall bar device}. \textbf{(a)} Longitudinal magnetoresistance for different angles $\theta$ between the applied magnetic field and the current direction. \textbf{(b)} Extracted Shubnikov-de Haas oscillations in the longitudinal resistance for the perpendicular field direction. The inset shows the FFT spectrum of the oscillations, with a clear peak at $F\approx$~17~T. \textbf{(c)} Oscillation frequencies from the Fourier transforms shown in the inset, as a function of the angle between the applied electric and magnetic fields. The fit to the data tells us that the oscillations originate from an elipsoidal Fermi surface with an anistropy of 3.3, which correpsonds to the bulk hole pocket.}
\label{fig:NL15}
\end{figure*}
Hall bar samples have been fabricated to characterize the Bi$_{0.97}$Sb$_{0.03}$ flakes used for this work. The Hall bars were fabricated by using standard e-beam lithography, followed by sputter deposition of Nb with a capping layer of a few nm of Pd. Measurements were performed at $T=$10~K so as to not induce superconductivity in the Bi$_{0.97}$Sb$_{0.03}$ flake.

The longitudinal magnetoresistance, as shown in figure~\ref{fig:NL15}(a), exhibits a rather sharp angle dependence, with a negative magnetoresistance of about 12\% for parallel electric and magnetic fields. Although subtle, Shubnikov-de Haas (SdH) oscillations can be observed. Figure~\ref{fig:NL15}(b) shows the SdH oscillations for perpendicular electric and magnetic fields, extracted by subtracting a simple smooth function from the measured data. The inset shows the obtained FFT spectrum, which reveals a single oscillation frequency $F\approx$17~T. The oscillation frequency shifts upon increasing the angle between the magnetic field and the $c$-axis of the crystal. Figure~\ref{fig:NL15}(c) shows the angle-dependence of this frequency, fitted with an ellipsoidal Fermi surface with an anisotropy of $k_z=3.3\cdot k_{x,y}$. This indicates that the oscillations originate from the bulk hole pocket~\cite{SChuan2018}. 

\section{Appendix C: Constant local electric field}
\begin{figure*}
\includegraphics[width=.96\textwidth]{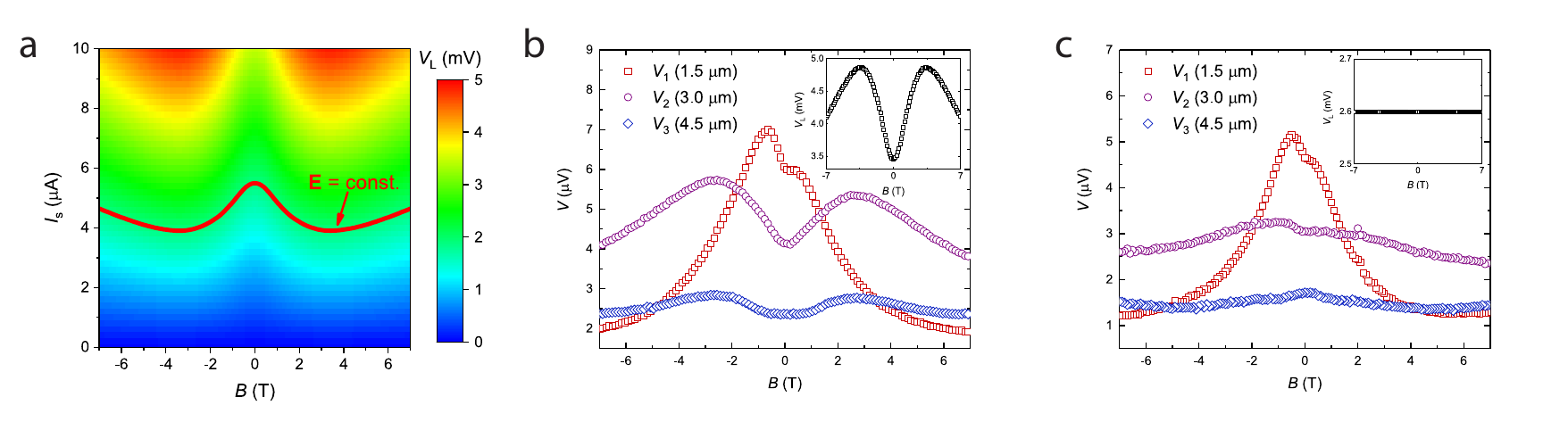}
\caption{\textbf{Extracting data for a constant electric field.} \textbf{(a)} Local voltage measured as a function of applied current and magnetic field. The red line corresponds to a constant local voltage for varying applied magnetic fields. \textbf{(b)} Measured non-local voltages when a constant current is applied at the source contacts and the local voltage changes with magnetic field as shown in the inset. \textbf{(c)} The extracted non-local data after following our method of retracing the red curve of panel (a) over each of the non-local voltage maps, thereby keeping the source voltage constant, as shown in the inset.}
\label{fig:Vcv}
\end{figure*}

When studying the CME, in the ideal case one measures the non-local response of the chiral anomaly as a function of the applied magnetic field only, i.e. keeping the applied electric field at the source contacts constant. However, due to the low resistance of our sample, we can not voltage bias our sample and must resort to a current source, thereby causing the current to be constant as a function of the applied magnetic field. When measuring the local electric field, we find that this field is dependent on the magnetic field as well. Hence, sweeping the magnetic field changes both the strength of $\textbf{E}$ and $\textbf{B}$, as follows directly from the chiral magnetic effect~\cite{SQLi2016}.

In order to obtain a data set with a constant electric field at the source contacts, we measured the local and non-local voltages as a function of both the applied current and magnetic field. The dependence of the local voltage on both parameters can be seen in figure~\ref{fig:Vcv}(a). From this map, we can find a line for which $V_{\text{L}}(I_s,B)$ is constant. This line is plotted on top of the map. By retracing the same line on the non-local voltage maps, we can extract the non-local voltages that correspond to the same constant local electric field and study the magnetic field dependence. In other words, we measure the local voltage $V_{\text{L}}(I_s,B)$, extract the source current $I_s(V_{\text{L}}=\text{cst.\,},B)$, and use this to find the non-local voltages for a constant local electric field: $V_{\text{NL}}(I_s(B),B)$.

\begin{figure*}
\includegraphics[width=.96\textwidth]{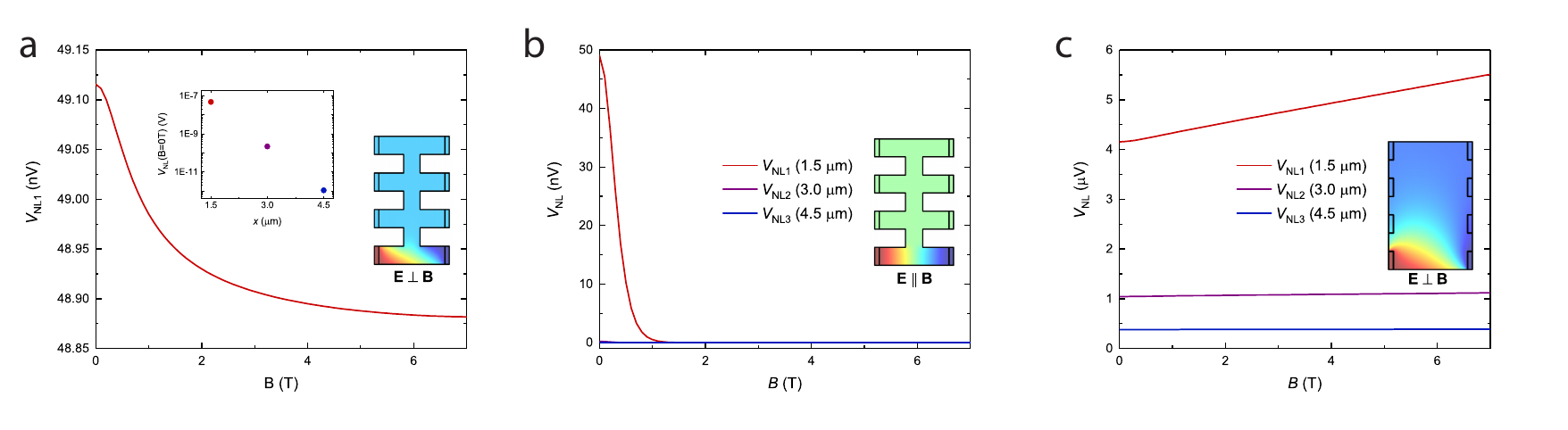}
\caption{\textbf{Simulation results for different geometries and/or field orientations.} For all panels, the device schematics show a contour plot of the voltage distribution for $B=1$ T. \textbf{(a)} Simulation results for a structured device in perpendicular magnetic fields. A small negative magnetoresistance is found at the nearest voltage probe, while for the other probes the signal is completely negligible. \textbf{(b)} Same as panel (a), but for magnetic fields parallel to the source voltage. Current jetting causes a large negative magnetoresistance at the nearest contact, but this signal has vanished for further contacts. \textbf{(c)} Same as panel (a), but for a non-structured device (i.e. a flake) in perpendicular magnetic fields. The negative magnetoresistance from panel (a) is no longer present, indicating that the negative magnetoresistance in perpendicular fields is a geometrical effect.}
\label{fig:COMSOL}
\end{figure*}

To increase the accuracy of the maps - we measured the field dependent data for 51 different values of the excitation current - we linearly interpolated our data as a function of $I_s$. Although recent work on Bi$_{1-x}$Sb$_x$ suggests that Ohm's law is violated in this system~\cite{SShin2017}, our generated electric field is well above the non-linear regime. 

Figure~\ref{fig:Vcv}(b) shows the raw data, i.e. when we omit this procedure and would measure the voltages by applying a constant current. We clearly observe that the local voltage is not constant with respect to the magnetic field. Figure~\ref{fig:Vcv}(c) shows the data when we follow our procedure. The local voltage is now constant as a function of the magnetic field and the large dips that were present in figure~\ref{fig:Vcv}(b) are less pronounced in figure \ref{fig:Vcv}(c). 

\section{Appendix D: Modelling the Ohmic contribution}
The measured data comprises a combination of a chiral signal and a normal, i.e. Ohmic, contribution. To gain better insight in the Ohmic contribution, we modelled this contribution in COMSOL Multiphysics. The device geometry used for the measurements was replicated in COMSOL. One of the source leads was defined as a current source, sourcing 1 $\mu$A, while the other current lead was set as ground. The other terminals were voltage probes. 

For simulations with perpendicular electric and magnetic fields (i.e. $B$ in the $z$-direction and $I$ in the $x$-direction), the three-dimensional conductivity tensor for a single band can be obtained from the Drude model~\cite{SDatta}; 
\begin{equation}
\hat{\sigma_\perp}=\sigma_0
\begin{pmatrix}
\frac{1}{1+(\mu B)^2} & \frac{\mu B}{1+(\mu B)^2} & 0 \\
-\frac{\mu B}{1+(\mu B)^2} & \frac{1}{1+(\mu B)^2} & 0 \\
0 & 0 & 1
\end{pmatrix},
\end{equation}
where $\sigma_0=ne\mu$. For parallel electric and magnetic fields (i.e. $B$ in the $x$-direction), one can show that the conductivity tensor is given by
\begin{equation}
\hat{\sigma_\parallel}=\sigma_0
\begin{pmatrix}
1 & 0 & 0 \\
0 & \frac{1}{1+(\mu B)^2} & \frac{\mu B}{1+(\mu B)^2} \\
0 & -\frac{\mu B}{1+(\mu B)^2} & \frac{1}{1+(\mu B)^2} \\
\end{pmatrix}.
\end{equation}
Upon changing the tensor in COMSOL we effectively rotate the magnetic field with respect to the sample. Figure~\ref{fig:COMSOL} shows the results of simulations for different structures (depicted as insets within the figures) and different orientations of the magnetic field with respect to the electric field. For every simulation we used a carrier density of $n=4\cdot10^{17}$~m$^{-2}$ and mobility $\mu=2$~m$^2$/Vs. 

Figure~\ref{fig:COMSOL}(a) shows the simulation results for a magnetic field perpendicular to the plane (and thus to the current). The sample is shaped into a non-local structure with a geometry as used in the experiments. We observe decreasing non-local voltages at the nearest contact with increasing perpendicular magnetic field. As the used model only captures the Drude conductivity, the observed decrease cannot be explained by a chiral component. The contacts further away from the source all returned negigibly small voltages.

\begin{figure*}
\includegraphics[width=.96\textwidth]{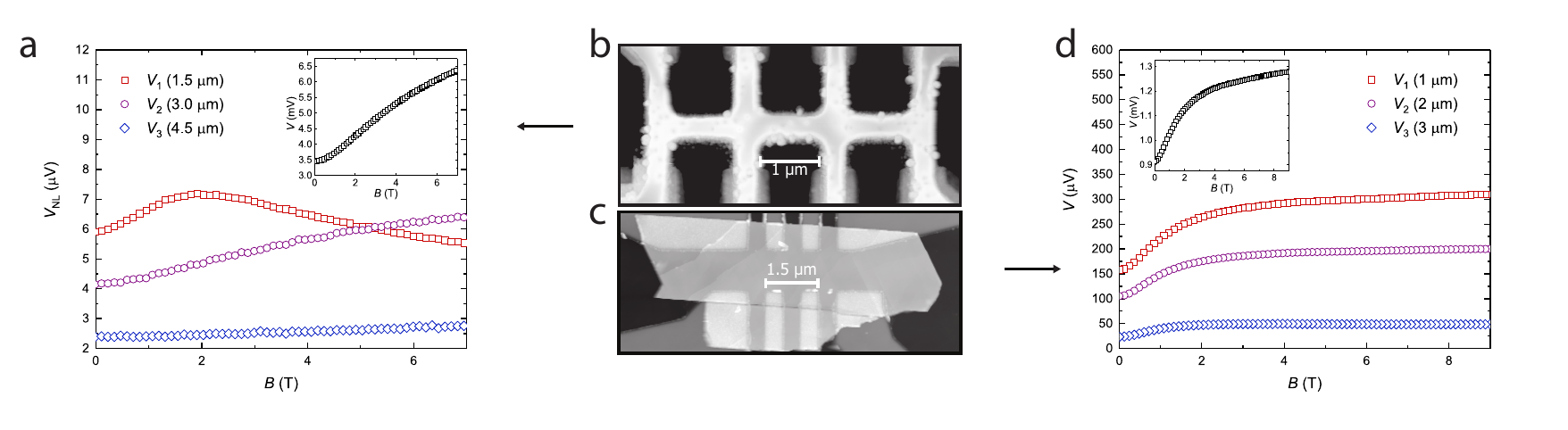}
\caption{\textbf{Measurement data for perpendicular electric and magnetic fields.} For all transport measurements in this figure, data has been measured for constant source current. \textbf{(a)} Magnetoresistance data measured for the non-local device of the main text, in perpendicular magnetic fields. Inset: voltage at the source contacts. \textbf{(b)} AFM image of the structured non-local sample as shown in the main text, but presented again for the ease of comparison. \textbf{(c)} AFM image of a non-local sample for which the flake has not been etched into a channelled structure. \textbf{(d)} Same as panel (a), but for the unstructured device.}
\label{fig:Perp}
\end{figure*} 

Figure~\ref{fig:COMSOL}(b) shows the non-local voltages for magnetic fields parallel to the electric source field. The strongly decreasing magnetoresistance can be explained by an effect called current jetting. Current jetting, as follows from the conductivity tensor for parallel fields, is an effect where the magnetic field suppresses current flow in the transverse direction~\cite{SPippard, Hirsch, dosReis}. This reduces the amount of current that spreads out towards the non-local voltage probes, resulting in a decrease in the voltages that are measured. The fact that current jetting can also introduce negative magnetoresistance, is (among other effects) one of the main reasons that observing negative magnetoresistance itself is not a proof of observing the chiral anomaly. However, the simulated non-local voltages decrease (even vanish) much faster than observed in experiments. From this, we conclude that the negative magnetoresistance in our samples is not caused by current jetting. 

In panel (a), we observed a decreasing non-local voltage while the electric and magnetic fields were perpendicular to each other. To further investigate the nature of this decrease, simulations were performed on a rectangular, unstructured flake with 8 contacts, as shown in figure.~\ref{fig:COMSOL}(c). The non-local voltages now show an upturn with increasing parallel magnetic field. By comparing panels (a) and (c), which make use of the same conductivity tensor and only differ in terms of geometry, we conclude that the decreasing non-local voltage in panel (a) is caused by the geometry of the device.


\begin{figure}
\includegraphics[width=.48\textwidth]{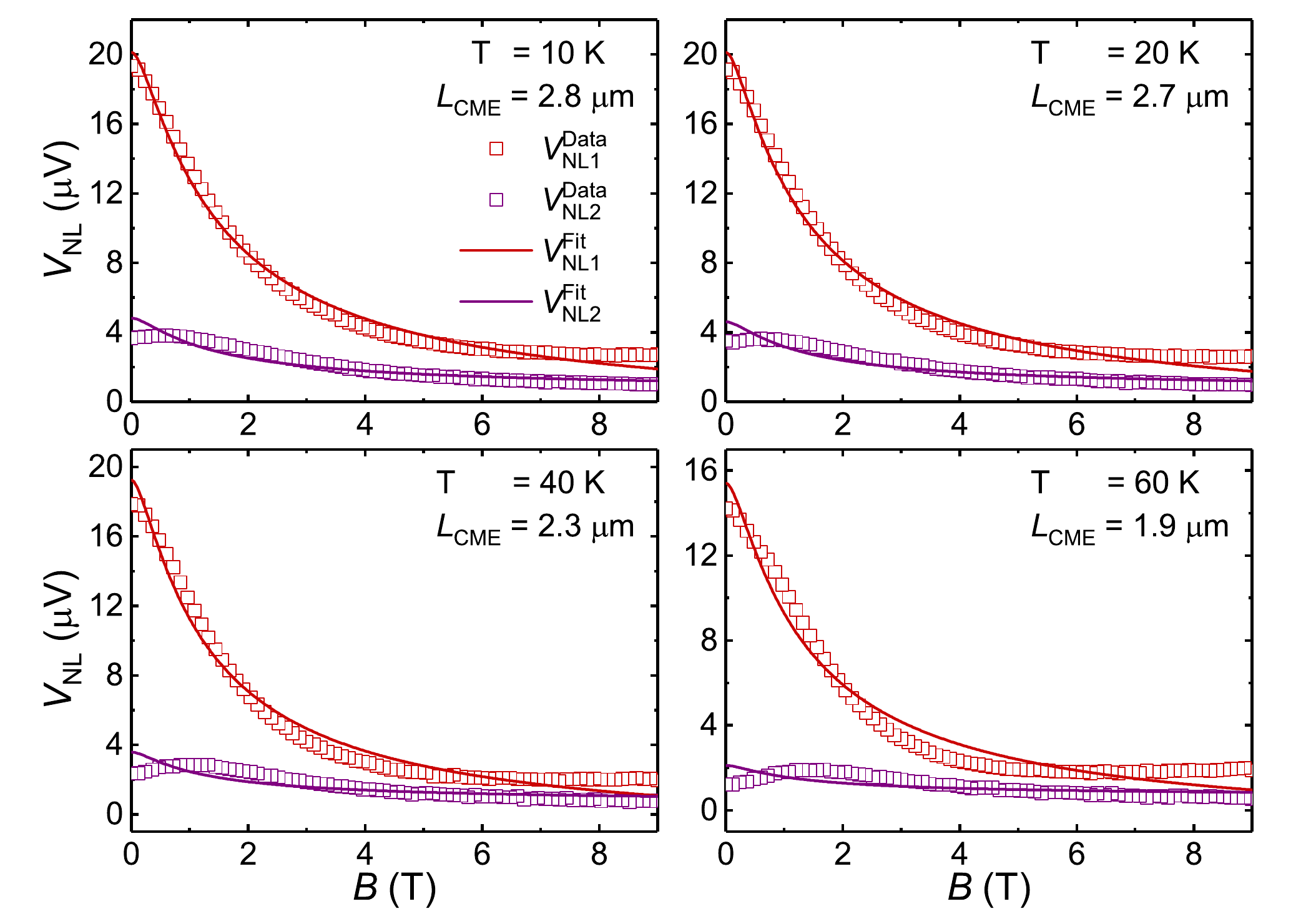}
\caption{\textbf{Fitting the intervalley diffusion length for different temperatures.} Examples of fits to the data in order to extract the chiral contribution (see main text) and Ohmic contribution (which is constant) simultaneously for the different temperatures.}
\label{fig:temp}
\end{figure}

\section{Appendix E: Non-local measurements at perpendicular electric and magnetic fields}
Measurements have been performed for perpendicular electric and magnetic fields. In figure~\ref{fig:Perp} we present the data for two different samples. Striking is that for panel (a) we observe a decrease  of $V_1$ as a function of magnetic field, while in the panel (d) we only observe the standard upturn with magnetic field, similar to the data measured for the Hall bar sample. 

In the simulations of the previous section, we found that this negative magnetoresistance in perpendicular fields is a geometrical effect. Comparing panels (a) and (b) of figure~\ref{fig:Perp} to panel (a) of the simulations (figure~\ref{fig:COMSOL}), and panels (c) and (d) of figure~\ref{fig:Perp} to panel (c) of the simulations, we see that the experimental data confirms this.
That the simulations are much more distinctive on the two different geometries, is because of the simulation taking only a single band into account for the conductivity, while in reality the conductivity is mediated by four channels. Furthermore, as can be seen from panel (c), the flakes are not always uniform in thickness and are not as well defined as the geometries in our simulations. 

\section{Appendix F: Temperature dependence of the chiral diffusion length}
The device used in the main text to study the temperature dependence had one broken voltage probe. As a consequence, only 2 non-local voltages were measured. This device has contacts spaced 1.5~$\mu$m apart, with twice this spacing in the middle, so that the contacts $V_{\text{NL1,2}}$ are located at 3.0~$\mu$m and 7.5~$\mu$m respectively. 

\begin{figure}
\includegraphics[width=.48\textwidth]{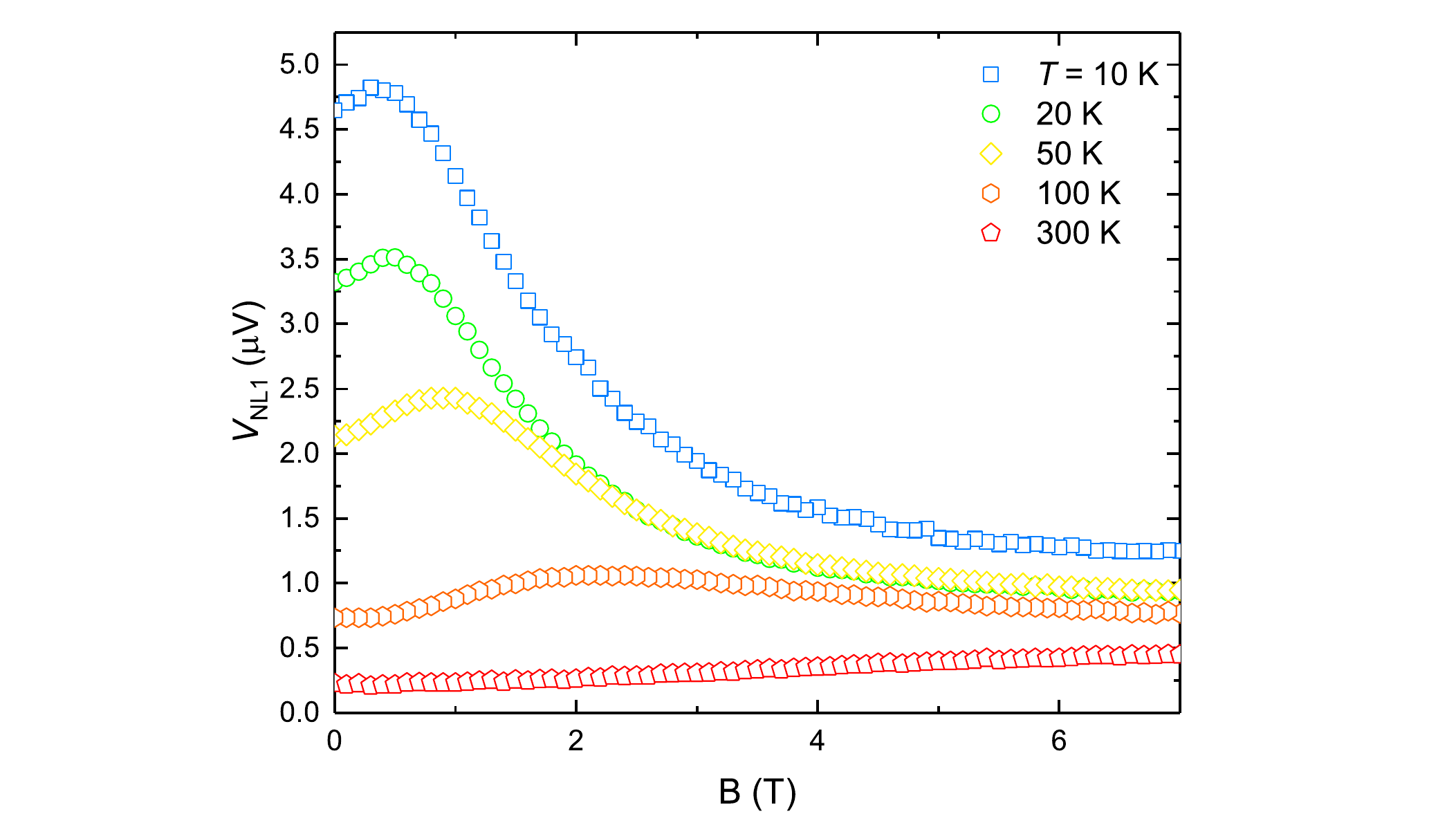}
\caption{\textbf{Temperature dependence of the smaller non-local device described in the main text.} The temperature dependence of the nearest non-local contact shows the same general trend as the device used for the temperature-dependence in the main text.}
\label{fig:temp2}
\end{figure}

The measurements for this device were performed using a single, constant current. Because of the linearity of the measured voltage as a function of the applied current (see section C), we can still extrapolate the data by using the recorded data point and the fact that $V_L(I_s=0)=0$~V. Then, the procedure as outlined in section C was followed to convert the mapped data into data for which the local electric field is effectively constant. In figure~\ref{fig:temp} we present some examples of fits to the data to extract the temperature-dependent chiral diffusion length as presented in the main text. As described in the main text, all fits were performed on the intermediate field data (2~T-5.5~T).


Lastly, figure \ref{fig:temp2} shows the temperature dependence of the non-local voltage of the smallest sample shown in the main text. Here, we also observe a decreasing strength of the anomaly with increasing temperature. The observed peak in the curve drifts to higher magnetic fields with temperature. For temperatures above 100 K, we do not observe negative magnetoresistance.



\begin{thebibliography}{99}
	
	\bibitem{Hsieh2008} D. Hsieh \textit{et al.}, 
	Nature \textbf{452}, 970 (2008).
	
	\bibitem{Kim2013} H.J. Kim \textit{et al.}, 
	Phys. Rev. Lett. \textbf{111}, 246603 (2013).
	
	\bibitem{Chuan2018}
	C.~Li, J.C.~de Boer, B.~de Ronde, S.V.~Ramankutty, E.~van Heumen, Y.~Huang, A.~de Visser, A.A.~Golubov, M.S.~Golden, A.~Brinkman,
	Nat. Mater. \textbf{17}, 875–880 (2018)
	
	\bibitem{Nagaosa2014}
	B.J.~Yang N.~Nagaosa,
	Nat. Commun. \textbf{5}, 4898 (2014) 
	
	\bibitem{Liang2018}
	S.~Liang, J.~Lin, S.~Kushwaha, J.~Xing, N.~Ni, R.J.~Cava, N.P.~Ong,
	Phys. Rev. X \textbf{8}, 031002 (2018)
	
	\bibitem{Parameswaran2014}
	S.A.~Parameswaran, T.~Grover, D.A.~Abanin, D.A.~Pesin, A.~Vishwanath,
	Phys. Rev. X \textbf{4}, 031035 (2014)
	
	\bibitem{Zhang2017}
	C.~Zhang, E.~Zhang, W.~Wang, Y.~Liu, Z.G.~Chen, S.~Lu, S.~Liang, J.~Cao, X.~Yuan, L.~Tang, Q.~Li, C.~Zhou, T.~Gu, Y.~Wu, J.~Zou, F.~Xiu,
	Nat. Commun. \textbf{8}, 13741 (2017)
	
	\bibitem{Moll2016}
	P.J.W.~Moll, N.L.~Nair, T.~Helm, A.C.~Potter, I.~Kimchi, A.~Vishwanath, J.G.~Analytis,
	Nature \textbf{535}, 266–270 (2016)
	
	\bibitem{Nielsen1983}
	H.B.~Nielsen, M.~Ninomiya
	Phys. Rev. Lett. B \textbf{130-6}, 389-396 (1983)

	\bibitem{SonSpivak2013}
	D.T.~Son, B.Z.~Spivak,
	Phys. Rev. B \textbf{88}, 1–4 (2013)
	
	\bibitem{Zyuzin2012}
	A.A.~Zyuzin and A.A.~Burkov
	Phys. Rev. B \textbf{86}, 115133 (2012)
	
	\bibitem{QLi2016}
    Q.~Li, D.E.~Kharzeev, C.~Zhang, Y.~Huang, I.~Pletikosić, A.V.~Fedorov, R.D.~Zhong, J.A.~Schneeloch, G.D.~Gu, T.~Valla,
    Nat. Phys. \textbf{12}, 550–554 (2016)
	
	\bibitem{HuiLi2013}
    Hui~Li, H.~He, H.Z.~Lu, H.~Zhang, H.~Liu, R.~Ma, Z.~Fan, S.Q.~Shen, J.~Wang
    Nat. Commun. \textbf{7}, 10301 (2016)
	
	\bibitem{Caizhen2015}
	C.Z.~Li, L.X.~Wang, H.~Liu, J.~Wang, Z.M.~Liao, D.P.~Yu,
	Nat. Commun. \textbf{6}, 10137 (2015)

	\bibitem{LiuAllen1995}
	Y.~Liu, R.E.~Allen,
	Phys. Rev. B \textbf{52-3}, 1566-1577 (2017)

\end{thebibliography}

\begin{thebibliography}{99}

	\bibitem{SNagaosa2014}
	B.J.~Yang N.~Nagaosa,
	Nat. Commun. \textbf{5}, 4898 (2014) 
	
	\bibitem{SZyuzin2012}
	A.A.~Zyuzin and A.A.~Burkov
	Phys. Rev. B \textbf{86}, 115133 (2012)
	
	\bibitem{SLiu2010}
	C.X.~Liu, X.L.~Qi, H.~Zhang, X~Dai, Z~Fang, S.C.~Zhang
	Phys. Rev. B \textbf{82}, 045122 (2010)
	
	\bibitem{STeo2008}
	J.C.Y.~Teo, L.~Fu, C.L.~Kane
	Phys. Rev. B \textbf{78}, 045426 (2008)
	
	\bibitem{SChuan2018}
	C.~Li, J.C.~de Boer, B.~de Ronde, S.V.~Ramankutty, E.~van Heumen, Y.~Huang, A.~de Visser, A.A.~Golubov, M.S.~Golden, A.~Brinkman,
	Nat. Mater. \textbf{17}, 875–880 (2018)
	
	\bibitem{SQLi2016}
    Q.~Li, D.E.~Kharzeev, C.~Zhang, Y.~Huang, I.~Pletikosić, A.V.~Fedorov, R.D.~Zhong, J.A.~Schneeloch, G.D.~Gu, T.~Valla,
    Nat. Phys. \textbf{12}, 550–554 (2016)
   	
	\bibitem{SShin2017} 
	D.~Shin, Y.~Lee, M.~Sasaki, Y.H.~Jeong, F.~Weickert, J.B.~Betts, H.J.~Kim, K.S.~Kim, J.~Kim
	Nat. Mater. \textbf{16}, 1096–1099 (2017).
	
	\bibitem{SDatta}
	S.~Datta, 
	Electronic Transport in Mesoscopic Systems, Cambridge University Press (1995)
	
	\bibitem{SPippard}
	A.B. Pippard,
	Magnetoresistance in metals, Cambridge University Press (1989).

	\bibitem{SHirsch},
	 M.~Hirschberger, S.~Kushwaha, Z.~Wang, Q.~Gibson, S.~Liang, C.A.~Belvin, B.A.~Bernevig, R.J.~Cava, N.P.~Ong,
	Nat. Mater. \textbf{15}, 1161–1165 (2016)

	\bibitem{SdosReis}
	R.D.~dos Reis, M.O~Ajeesh, N.~Kumar, F.~Arnold, C.~Shekhar, M.~Naumann, M.~Schmidt, M.~Nicklas, E.~Hassinger,
	New Journ. of Phys. \textbf{18} (2016)
	
	\bibitem{SParameswaran2014}
	S.A.~Parameswaran, T.~Grover, D.A.~Abanin, D.A.~Pesin, A.~Vishwanath,
	Phys. Rev. X \textbf{4}, 031035 (2014)
	
	\bibitem{SZhang2017}
	C.~Zhang, E.~Zhang, W.~Wang, Y.~Liu, Z.G.~Chen, S.~Lu, S.~Liang, J.~Cao, X.~Yuan, L.~Tang, Q.~Li, C.~Zhou, T.~Gu, Y.~Wu, J.~Zou, F,~Xiu,
	Nat. Commun. \textbf{8}, 13741 (2017)
	
	\bibitem{SLiang2018}
	S.~Liang, J.~Lin, S.~Kushwaha, J.~Xing, N.~Ni, R.J.~Cava, N.P.~Ong,
	Phys. Rev. X \textbf{8}, 031002 (2018)

\end{thebibliography}
\end{document}